%% file: paper.tex
\documentclass[manuscript,screen,nonacm]{acmart}

\pdfoutput=1

\input{preamble/preamble}

\input{preamble/variables}

\copyrightyear{2021}
\acmYear{2021}
\acmConference[]{}{}
\acmBooktitle{}
\acmPrice{}
\acmDOI{10.1145/xxxx.xxxx}
\acmISBN{978-1-4503-8312-7/21/04}

\author{Manolis Chalkiadakis}
\affiliation{
	\institution{FORTH/University of Crete}
	\country{Greece}
}
\author{Alexandros Kornilakis}
\affiliation{
	\institution{FORTH/University of Crete}
	\country{Greece}
}
\author{Panagiotis Papadopoulos}
    \affiliation{
	\institution{Telefonica Research}
	\country{Spain}
}
\author{Evangelos P. Markatos}
\affiliation{
	\institution{FORTH/University of Crete}
	\country{Greece}
}
\author{Nicolas Kourtellis}
\affiliation{
	\institution{Telefonica Research}
	\country{Spain}
}

\begin{document}
\title{The Rise and Fall of Fake News sites: A Traffic Analysis}
\pagestyle{plain} 

\input{sections/0_abstract}

\maketitle

\input{sections/1_introduction}
\input{sections/2_data_collection}
\input{sections/3_traffic_analysis}
\input{sections/4_longitudinal}
\input{sections/5_ml-classification}
\input{sections/6_related}
\input{sections/7_conclusion}
\input{sections/8_acks}

\bibliographystyle{unsrt}
\balance
\bibliography{reference}
\end{document}

%% file: preamble/preamble.tex
\usepackage{booktabs}
\usepackage{pbox}
\usepackage{epsfig,endnotes}
\usepackage{epstopdf}
\usepackage{xcolor}
\usepackage{graphicx}
\usepackage{xspace}
\usepackage{enumitem}
\usepackage{url}
\usepackage{balance}
\usepackage{subcaption}
\usepackage{breakurl}
\usepackage{pbox}
\usepackage{times}  
\usepackage{pdfrender}
\usepackage{csquotes}
\usepackage[font=small,labelfont=bf]{caption}

\newcommand{\point}[1]{\par\smallskip\noindent\textbf{#1}. }

%% file: preamble/variables.tex
\newcommand{\totalBiased}{834\xspace}
\newcommand{\totalFake}{283\xspace}
\newcommand{\accuracy}{F1 score up to 0.942 and AUC of ROC up to 0.976\xspace}

\newcommand{\etc}{etc.}
\newcommand{\eg}{e.g., }
\newcommand{\ie}{i.e., }
\newcommand{\etal}{et al. }

%% file: sections/0_abstract.tex
\begin{abstract}

    Over the past decade, we have witnessed the rise of misinformation on the Internet, with online users constantly falling victims of fake news. A multitude of past studies have analyzed fake news diffusion mechanics and detection and mitigation techniques. However, there are still open questions about their operational behavior such as: How old are fake news websites? Do they typically stay online for long periods of time? Do such websites synchronize with each other their up and down time? Do they share similar content through time? Which third-parties support their operations? How much user traffic do they attract, in comparison to mainstream or real news websites? In this paper, we perform a first of its kind investigation to answer such questions regarding the online presence of fake news websites and characterize their behavior in comparison to real news websites. Based on our findings, we build a content-agnostic ML classifier for automatic detection of fake news websites (\ie \accuracy) that are not yet included in manually curated blacklists.
\end{abstract}

%% file: sections/1_introduction.tex
\section{Introduction}
Lots of things we read online may appear to be true, often is not. False information is news, stories or hoaxes created to deliberately misinform or deceive readers. Usually, these stories are created to (i) either lure users and become a profitable business for the publishers (\ie clickbait) or (ii) influence people’s views, push a political agenda or cause confusion. False information can deceive people by looking like trusted websites or using similar names and web addresses to reputable news organisations. 

The most important types of fake news include:
(a) \emph{Clickbait}: stories that are carefully fabricated to gain more website visitors and drive advertising revenues for publishers. Such stories use sensationalist headlines to grab attention and increase Click-through rates normally at the expense of truth or accuracy. (b) \emph{Propaganda}: stories that are created to deliberately mislead audiences, promote a biased point of view or particular political agenda. (c) \emph{Sloppy Journalism}: stories with unreliable information or without verified facts which can mislead audiences. (d) \emph{Misleading Headings}: stories that are not completely false but distorted using misleading or sensationalist headlines, in such a way that can spread quickly via social media sites where only headlines and small snippets of the full article are displayed on audience newsfeeds. (e) \emph{Satire}: stories for entertainment and parody (\eg the Onion, The Daily Mash, \etc).

An analysis~\cite{fakeNewsVSrealNews} found that on Facebook, the top 20 fake news stories about the 2016 U.S. presidential election received more engagement than the top 20 election stories from 19 major media outlets.
According to a different study~\cite{fakeNewsSurvey}, US citizens rate fake news as a larger problem than racism, climate change, or terrorism. According to the study, more than making people believe false things, the rise of fake news is making it harder for people to recognize the truth, thus, making them especially conservative and less well-informed. 

Considering the importance of this emerging threat to society, there is a significant body of research in the last years, aiming to analyze the content, the methodologies, the possible detection and mitigation techniques or the way fake news spread (\eg ~\cite{zannettou2019web,Ghanem2019AnEA, 7589045,10.1145/3131365.3131390,Jin2017RumorDO,10.1145/3305260, zannettou2019disinformation}).
In fact, fake news involved in the 2016 elections has received significant attention and well studied (\eg~\cite{grinberg2019fake}), although several follow a more generic approach of analysis~\cite{lazer2018science} .
There have also been works on the spread of fake news on social networks.
For example, Shao \etal~\cite{shao2017spread} studied the spread of fake news by social bots.
Also, Fourney \etal~\cite{fourney2017geographic} conducted a traffic analysis of websites known for publishing fake news in the months preceding the 2016 US presidential election.

Apart from the important existing works in the area, yet little we know about the network characteristics of fake news distributing websites:
What is the lifetime of these websites?
What is the volume of traffic they receive and how engaged is their audience?
How do they connect with other marked-as-fake news sites?
In this study, we take the first step in answering these questions.
We collect a dataset of \totalFake websites tagged from known fact-checking lists as delivering fake news and perform traffic and network analysis of such websites.
In particular, and contrary to related work (\eg~\cite{fourney2017geographic}), we study and compare the user engagement of fake and real news sites by analyzing traffic-related metrics. Additionally, we explore the lifetime of fake news sites and their typical uptime periods over a time range of more than 20 years, and we propose a methodology to detect websites that are synchronizing not only their uptime periods but also the content they serve during these periods. Based on our findings, we design a content-agnostic ML classifier for the automatic detection of fake news websites.

\noindent{\bf Contributions} In summary, this paper makes the following main contributions:
\begin{enumerate}[label=(\roman*),leftmargin=0.5cm, topsep=-0.3cm]
\item We conduct the first of its kind temporal and traffic analysis of the network characteristics of fake new sites aiming to shed light on the user engagement, lifetime and operation of these special purpose websites. We compose an annotated dataset of \totalFake fake news sites indicating when such websites are alive or dormant, which we provide open sourced\footnote{\url{https://github.com/mxalk/fake_news_resources}}.

\item We propose a methodology to study how websites may be synchronizing their alive periods, and even serving the exact same content for months at a time. We detect numerous clusters of websites synchronizing their uptime and content for long periods of time within the USA presidential election years 2016-2017.

\item We study the third party websites embedded in the different types of news sites (real and fake) and we find that during the aforementioned election years there is a significant increase in the use of analytics in the fake news sites but not an increase in the use of ad related third-parties. Additionally, domains like \textit{doubleblick}, \textit{googleadservices} and \textit{scorecardresearch} tend to have higher presence in real news sites than in fake ones. On the contrary, \textit{facebook} and \textit{quantserve} have higher presence in fake news sites.

\item We build a novel, content-agnostic ML classifier for automatic detection of fake news websites that are not yet included in manually curated blacklists. We tested various supervised and unsupervised ML methods for our classifier which achieved \accuracy.
\end{enumerate}

%% file: sections/2_data_collection.tex
\section{Data Collection}
\label{sec:dataset}

To perform this study, we collect data from different sources and in this section we describe in detail our data.
First, we obtain lists with manually curated news sites, categorized as ``fake'' and ``real''.
We then use the ``fake'' news sites list as input for crawling historical data from Wayback machine to annotate the state of each website.
Finally, to explore the web traffic characteristics of the two categories of news sites and how their audience behaves, we collect data from SimilarWeb~\cite{similarweb} and CheckPageRank~\cite{checkpagerank}.

\subsection{Fake and Real News Sites Dataset}
For this study, we compose two manually curated lists of news sites. One with sites that are marked as ``fake'' and one with sites marked as ``real''.
For the fake news sites, we utilize the domains repository provided by the {\texttt opensources.co} website~\cite{OpenSources2017}. The repository contains \totalBiased biased news sites, of which \totalFake domains are manually checked and flagged as ``fake''.
This is a well-accepted list, and has been used in studies related to the fake news ecosystem of 2016 US elections~\cite{bovet2019influence,horne2018assessing}, as well as in fake news detection tools such as the $BS$-$Detector$~\cite{bsdetector}.
Additionally, we compose a second list of same size, for ``real'' news sites by taking the top Alexa news sites (and ensuring that there are no sites there that are marked as fake in the $opensources$ repository).

\subsection{Web Traffic and Audience Behavioral Analytics}
\label{sec:traffic-data}
To assess the user engagement in the websites of our dataset, we collect web traffic data from popular data services like SimilarWeb and CheckPageRank (date of crawl: June'20).
SimilarWeb provides Web and other traffic related data per website, while CheckPageRank provides search engine-related information.
In summary, we analyze volume of user visits, where the user visits come from, their duration and what subdomains they browse, the number of users who bounce off a domain, as well as Web connectivity of websites with respect to number and type of incoming or outgoing links from and to other sites.

\subsection{Historical Data \& Annotation of State}
Next, we focus on the \totalBiased news sites flagged for spreading misinformation (\ie marked as fake or biased) and to identify their different states across time, we collect historical data from the Wayback Machine~\cite{wayback}. Specifically, we first query the Wayback CDX server for each such news site in our list, and we get an index of the available timestamps for the particular domain. Then, we proceed with downloading the landing page of each timestamp and storing it locally for further processing. In total, we downloaded the content of these websites from the last 23 years.

\begin{figure}[t]
	    \centering		
	    \includegraphics[width=0.45\linewidth]{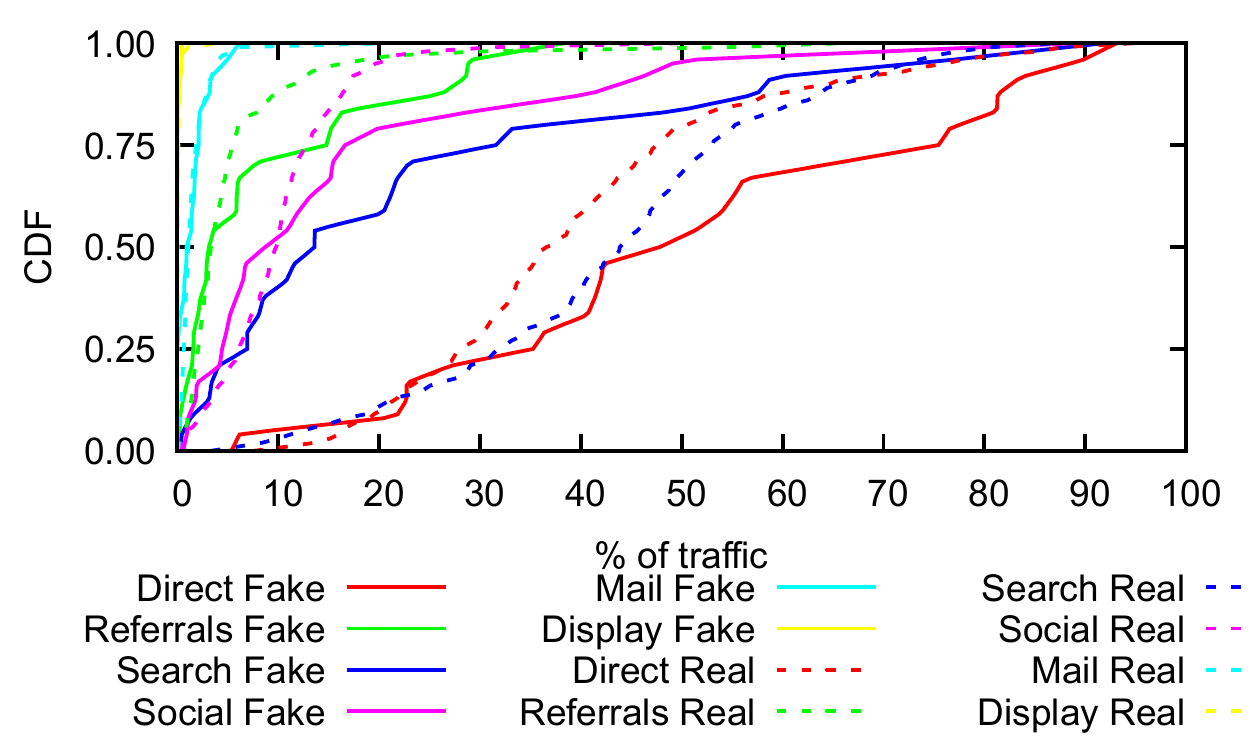}\vspace{-0.3cm}
	    \caption{CDF of traffic sources for real and fake news sites. The median fake news site is being accessed mostly directly, when the median real news site is being accessed mostly via search engines.}\vspace{-0.3cm}
	    \label{fig:similarweb_traffic_sources}
\end{figure}

The landing pages of each timestamp reflect the state at which the websites were in during that timestamp. The pages could contain material related to the domain crawled, or irrelevant content: if the domain name is not paid and is returned to the market for sale. To understand what is the state of each website for each timestamp, we attempt a manual annotation for each snapshot. By using the Puppeteer~\cite{puppeteer} framework, we iterate through the timestamped versions of these domains and render them on screen. In each case, a dialogue box is prompted, and we categorize the timestamped website as one of the following:

\begin{enumerate}[label=(\roman*)]
    \item \textbf{alive}: We consider a website being ``alive'' when it is offering news content
    \item \textbf{zombie}: We consider a website being ``zombie'' when it is offering content other than news (\eg e-marketing or other news-irrelevant content).
    \item \textbf{dead}: If the timestamped content of a website is none of the above (\eg no HTML content was returned, or HTTP errors were returned), the website is declared ``dead''.
\end{enumerate}

Given that a website may have been archived multiple times per month on Wayback, we aggregate the state of the website per month, by making the assumption that if it had at least one ``alive'' timestamp in said month, then it was ``alive'' for the entire month. Similarly, it was in a ``zombie'' state, if it had at least one such state in that month, or ``dead'', if none of the above applied.
Finally, if there is a month that Wayback does not have a state for a website, that timestamp is marked as ``missing'' for said website.

%% file: sections/3_traffic_analysis.tex
\section{User Engagement}
\label{sec:traffic}

As a first step, we set out to analyze and compare various user traffic-related metrics for the fake and real news sites in our lists, in an attempt to understand what is the different behavior of the audience in these two categories of websites.
In particular, we focus on (1) where users come from to land on such websites, (2) how many pages they visit within a website, (3) their visit's duration and what sub-domains they browse, (4) the number of users who bounce off a domain, and (5) the Web connectivity of websites with respect to number and type of incoming or outgoing links to other sites.

\noindent{\bf Where do users come from?}
In Figure~\ref{fig:similarweb_traffic_sources}, we study the different sources that drive traffic to fake and real news sites. As we can see, the median fake site is being accessed mostly  directly (user navigates directly to the website) by the users ($Direct$), or via links in $Social$ media and search engines ($Search$). On the other hand, the median real news site is being accessed mostly via search engines ($Search$), and $Direct$ follows.
Sources such as $Mail$ and $Display$ drive similar traffic to fake and real news sites.

\noindent{\bf How many pages do users visit?}
In Table~\ref{tbl:stats}, we present the mean, standard deviation, median and 90th percentile of our user engagement metrics across all fake and real news sites.
If we focus on the average number of pages per visit (in a time window of 6 months), we see that the median real news site tends to have a larger number of pages (\ie 2.18 pages, on average) visited per user than the median fake news site (\ie 1.72 pages, on average).
Considering the 90th percentile, however, we see that there are fake news sites that have more (up to 3.54) pages visited on average than the corresponding real news sites (\ie 3.49 pages visited, on average).

\begin{figure}[t]
\centering
    \begin{minipage}[t]{0.48\textwidth}
        \centering
		\includegraphics[width=1.0\linewidth]{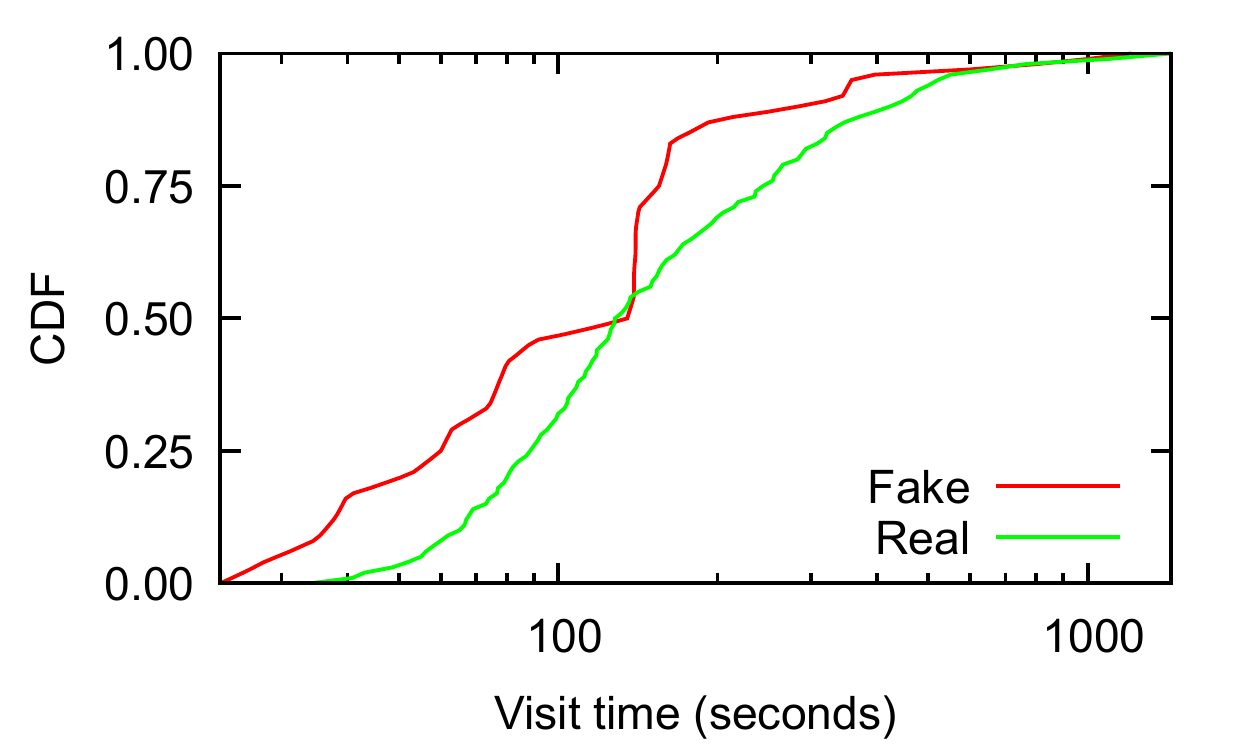}
		\vspace{-0.6cm}
		\caption{Distribution of visit duration in seconds, for real and fake news sites. In real news sites, the visit duration is longer than in fake news sites and follows a distribution close to power law.}\vspace{-0.3cm}
		\label{fig:similarweb_visit_duration}
    \end{minipage}
    \hfill
	\begin{minipage}[t]{0.48\textwidth}
	    \centering
		\includegraphics[width=1.0\linewidth]{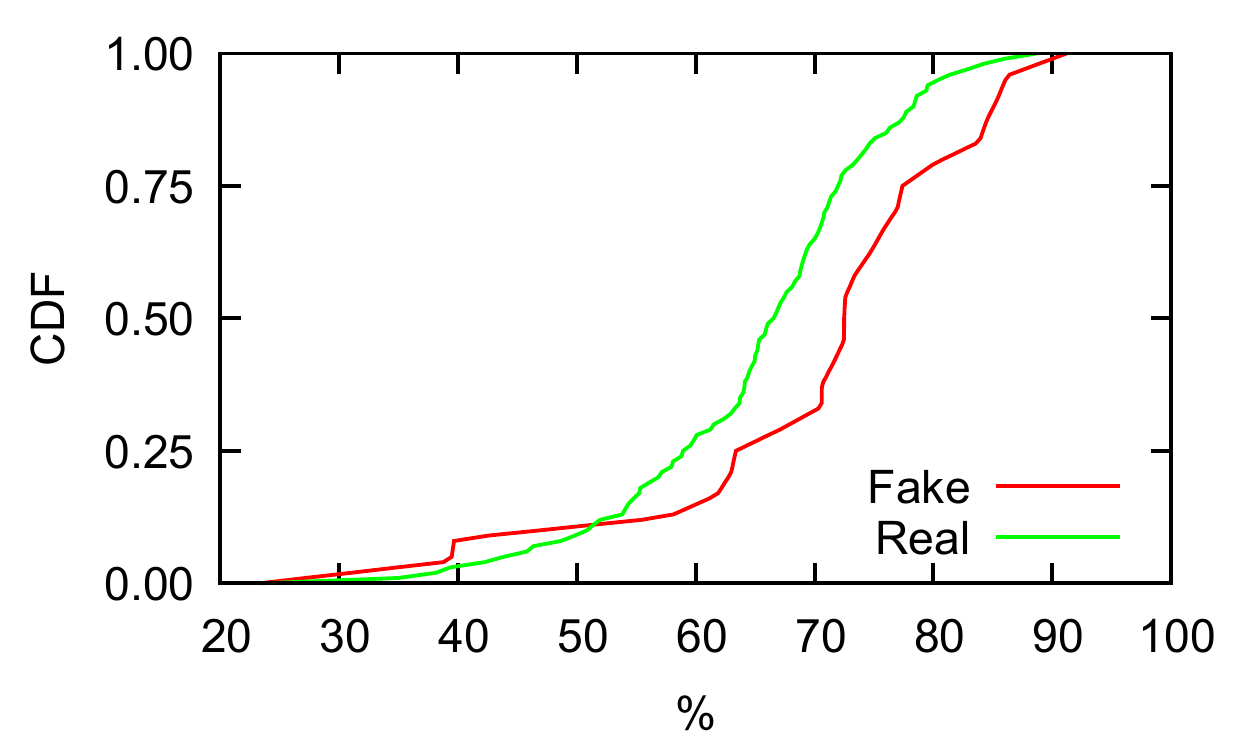}
		\vspace{-0.6cm}
	    \caption{Distribution of bounce rate for real and fake news sites. The median real news site has a lower bounce rate (\ie 66.55\%), compared to the median fake news one (\ie 72.49\%).}\vspace{-0.3cm}
	    \label{fig:similarweb_bounce_rate}
	\end{minipage}\vspace{-0.3cm}
\end{figure}

\noindent{\bf How long do users stay per visit?}
In Figure~\ref{fig:similarweb_visit_duration}, we present the distribution of the average duration of the user visits per news site. This duration defines the time elapsed between the beginning of the first and the end of last page visit (sessions are considered closed after 30 minutes of user inactivity~\cite{similarweb-average-visit}). As we can observe, in real news sites, the visit duration follows a distribution close to power law. Additionally, as presented also in Table~\ref{tbl:stats}, visits last longer (\ie 198.8 seconds) in real news sites, than in fake news sites (\ie 163.4 seconds), with the visits of the 90th percentile lasting around 423.40 seconds in real news sites and 284.20 seconds in fake news sites, on average.

\point{Website Bounce Rate}
In Figure~\ref{fig:similarweb_bounce_rate}, we present the percentage of visitors who enter a site and then leave after visiting only the first page (also known as bounce rate).
This metric is calculated by dividing the single-page sessions by all sessions~\cite{similarweb-bouncerate}, and reflects how well a site is doing at retaining its visitors.
A very high bounce rate is generally a warning that people are not willing to stick around to explore the website, and instead they choose to leave.
As we can see in the figure, the median real news site has a significantly lower bounce rate (\ie 66.55\%), compared to the median fake news one (\ie 72.49\%) with the corresponding rates for the 90th percentile being at 78.33\% and 85.08\%, respectively.
We deduct that fake news sites probably provide content of lower quality, that is less engaging or interesting compared to the real news sites.

\begin{figure}[t]
	\centering
	\begin{minipage}[t]{0.48\textwidth}
	\centering
	    \includegraphics[width=1.05\linewidth]{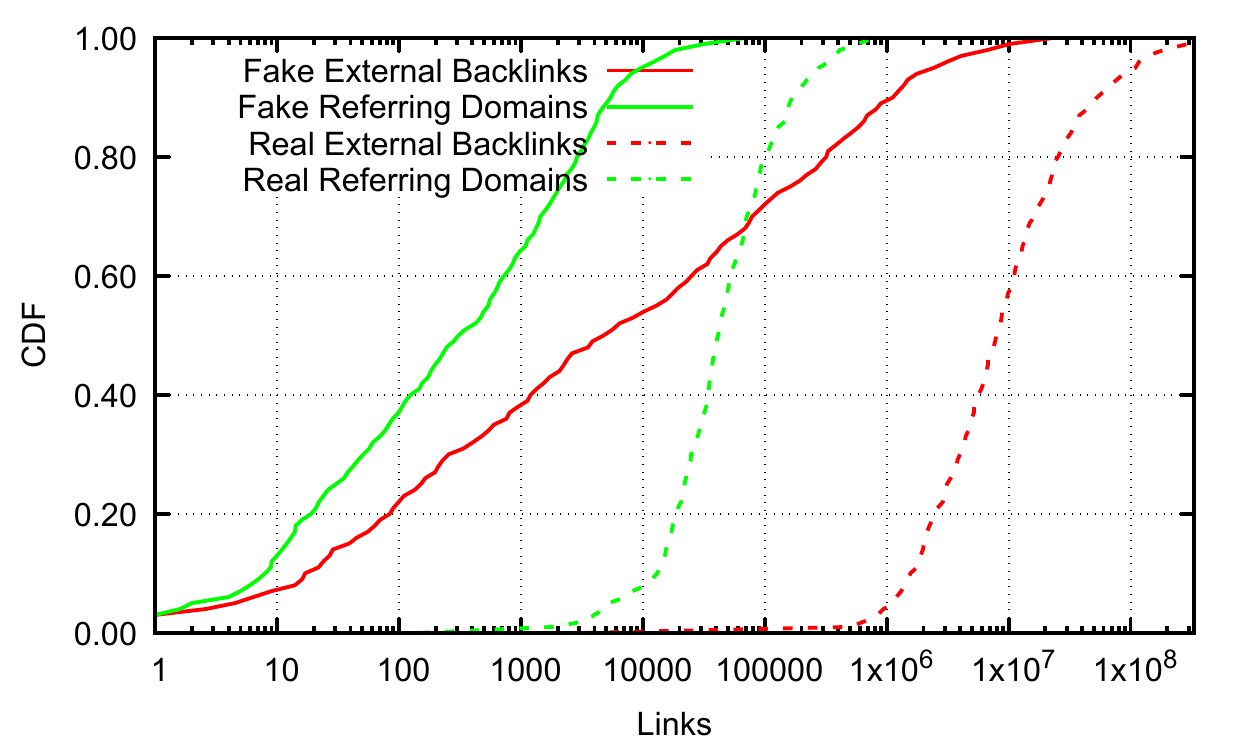}\vspace{-0.3cm}
    \caption{Distribution of backlinks and referring domains, for real and fake news sites. 
    In median values, they are significantly lower (4.7K backlinks) compared to real news (23.2M backlinks). Additionally, fake news sites have about 2 orders of magnitude lower (\ie 307) number of referring domains than real news (\ie 41.3K).}
    \label{fig:checkpagerank_links_cdf}
    \end{minipage}
    \hfill
	\begin{minipage}[t]{0.48\textwidth}
	\centering
		\includegraphics[width=1.05\linewidth]{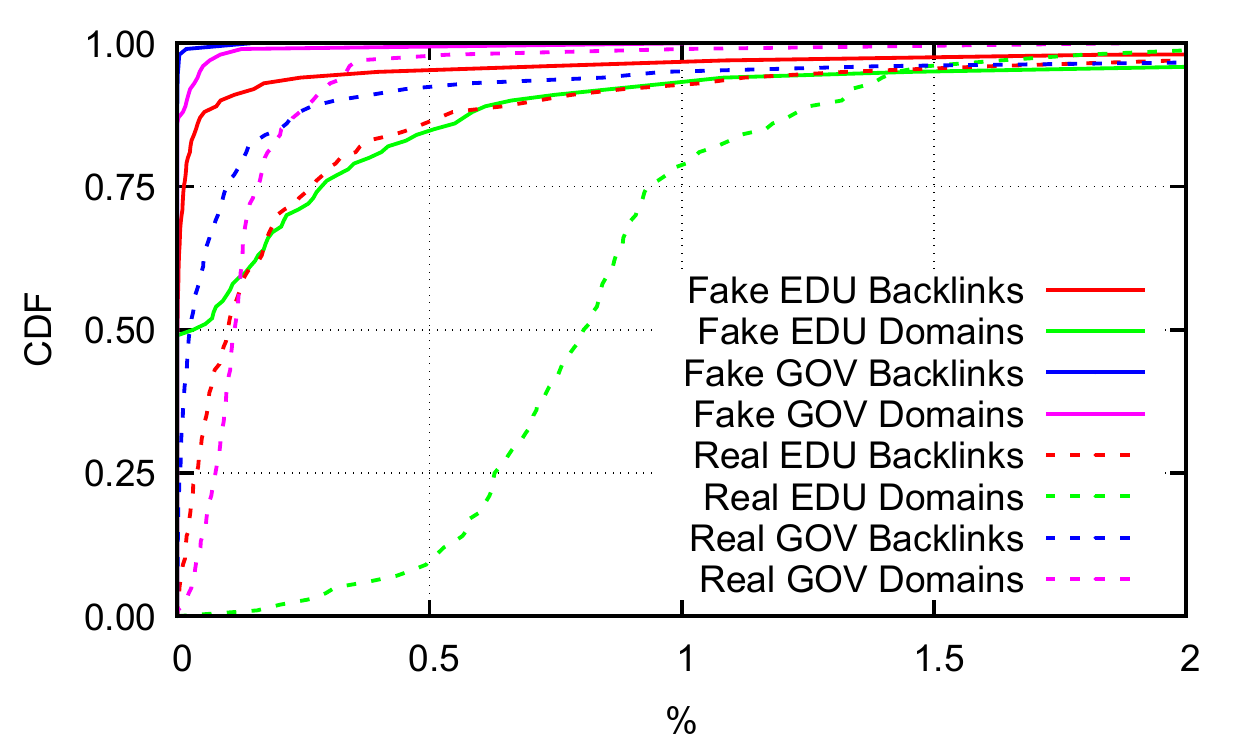}\vspace{-0.3cm}
    \caption{Distribution of EDU/GOV backlinks \& referring domains over total, for real and fake news sites. Fake news sites have lower portions of EDU backlinks and referrals, as well as GOV backlinks than the real news sites.}
    \label{fig:checkpagerank_links_ratios_cdf}
    \end{minipage}\vspace{-0.3cm}
\end{figure}

\point{Website Backlinks \& Referrals}
In Figure~\ref{fig:checkpagerank_links_cdf}, we draw the distribution of the number of backlinks for fake and real news sites.
A backlink (also called citation or inbound/incoming link) of a website $A$ is a link from some other website $B$ (\ie the referrer) to website $A$ (\ie the referent).
As we can see in the figure, backlinks of fake news sites follow a power law distribution and they are significantly lower compared to the backlinks of real news sites.
In particular, the median fake news site in our dataset scores  4.7K backlinks when the median real news site scores 23.2M backlinks! The 90th percentile of fake news sites has 1.12M backlinks when the corresponding real news site scores as high as 53M backlinks. It is of no doubt that this difference is caused by the lack of trust that a large portion of websites show to fake news distributing websites.

\begin{table}[t]
\centering
\footnotesize
\begin{tabular}{llrrrr} 
 	\toprule
	\textbf{Metric}	&	\textbf{Category}   & \textbf{Mean} & \textbf{Standard Deviation}  & \textbf{Median}    & \textbf{90th Percentile} \\
	\midrule
	Rank		&	Fake       & 197K  & 182K      & 130K      & 451K \\ 
			&	Real       & 13.8K & 16.6K     & 7.5K      & 33.7K \\
	\midrule
	Total Visits	&	Fake       & 3.3M  & 13M       & 349K      & 2.4M \\ 
				&	Real       & 44.5M & 115M      & 9M        & 92.2M \\
	\midrule
	Pages per Visit	&	Fake       & 2.33  & 2.14      & 1.72      & 3.54 \\ 
				&	Real       & 2.49  & 1.65      & 2.18      & 3.49 \\ 
	\midrule
	Visit Duration (in seconds)	&	Fake       & 163.4 & 234.3     & 135       & 284.2 \\ 
						&	Real       & 198.8 & 190.7     & 128       & 423.4 \\  
	\midrule
	Bounce Rate			&	Fake       & 69.42\%  & 15.81\%      & 72.49\%       & 85.08\% \\ 
						&	Real       & 65.13\%   & 11.04\%      & 66.55\%     & 78.33\% \\ 
	\midrule
	Backlinks				&	Fake       & 519K  & 1.94M     & 4.7K      & 1.12M \\ 
						&	Real       & 23.2M & 4.73M     & 23.2M     & 53M \\ 
	\midrule
	Referring Domains		&	Fake       & 2.2K  & 5.6K      & 307       & 5.4K \\ 
						&	Real       & 78K   & 106K      & 41.3K     & 168K \\ 
	\bottomrule
\end{tabular}\vspace{0.2cm}
	\caption{Summary of User engagement metrics for the two categories of websites in our dataset: Pages and Duration per Visit, Bounce rate, Backlinks and Referring Domains statistics.}\vspace{-0.5cm}
\label{tbl:stats}
\end{table}

Similarly, in Figure~\ref{fig:checkpagerank_links_cdf}, we plot the distribution of the number of referring domains per website. When backlinks are the links on the websites that link back to a given site, a referring domain is where backlinks are coming from\footnote{Example: think of the referring domain as a phone number and backlinks as the number of times you've gotten a call from that particular number.}. As we can see in the figure, in median values fake news sites have about 2 orders of magnitude lower (\ie 307) number of referring domains than real news (\ie 41.3K).

Finally, we study a particular class of domains or links from EDU or GOV domains, which could provide more authority and trust to a website when being referenced or linked to. In Figure~\ref{fig:checkpagerank_links_ratios_cdf}, we plot the portion of backlinks and referring domains related to EDU/GOV domains for fake and real news sites. We see that fake news sites have clearly lower portions of EDU backlinks and referrals, as well as GOV backlinks than the real news sites.

%% file: sections/4_longitudinal.tex
\section{Longitudinal Study of Fake News sites}

In this section, we focus on the fake news ecosystem and perform a historical analysis by studying the following questions:
(1) What is the lifetime of a fake news site?
(2) Are there any such websites that synchronize their uptime and reproduce the same content through time?
(3) Which third-party trackers were persistently embedded in such websites through time?
\subsection{What is the lifetime of a fake news site?}
\label{sec:lifetime}
We use three terms to study the lifetime of a fake news site.
First, we define with ``lifespan'' the upper limit for which a website may have existed on the Web.
This is computed as the time difference from the first and last timestamp with ``alive'' state.
Furthermore, the terms ``alive time'' and ``zombie time'' define the number of timestamps (\eg months) that the website under study has been tagged as ``alive'' or ``zombie'', respectively. Consequently, the timestamps for which the Wayback does not provide any data are considered ``dead''.

During the lifetime of a website, various problems could arise, such as the owner not paying for the domain for some months, or the website being offline due to some technical issues, \etc During such periods, and due to the crawling nature of the Wayback Machine, not all websites are archived at the same rate, and therefore, we may not have snapshots of websites for all timestamps studied.
In an attempt to infer what the state of a website was in such un-archived or ``missing'' timestamps, we use a 2-phase interpolation process.

\point{Phase 1} In the first phase ($P1$), we identify for each website any gaps between two timestamps with the same label $A$ ($A$=$\{alive, zombie\}$).
Thus, when the two timestamps are non-consecutive, and there is no other labelled timestamp between them, we proceed with propagating label $A$ to all ``missing'' timestamps of that gap.
For example, if website $W$ was found \textit{alive} in timestamps $i$ and $j$, with $m$ timestamps in-between them (\ie $j=i+m$), and no other state was captured between $i$ and $j$ (\ie during the $m$ timestamps), then, we assume that $W$ was alive for all the timestamps between $i$ and $j$.
Similar process was applied if $A$ was \textit{zombie}.
This interpolation process can be applied for increasingly larger gaps, \ie, for $m=1, 2,\dots$.
Therefore, we applied it for increasingly larger $m$, and we stopped at 3-year gaps (\ie m=36), since after that time there was no more correction done to the dataset.
\point{Phase 2}
In the second phase ($P2$), we identify gaps on the output of P1 between two ``alive'' timestamps up to three years apart, but allowing for up to 12 ``non-alive'' (\ie ``zombie'' or ``dead'') timestamps between them. The ``missing'' timestamps between these two alive ones were also labelled as ``alive''.

\begin{figure}[t]
	\centering
	\begin{minipage}[t]{0.48\textwidth}
	    \centering
	    \includegraphics[width=1.05\linewidth]{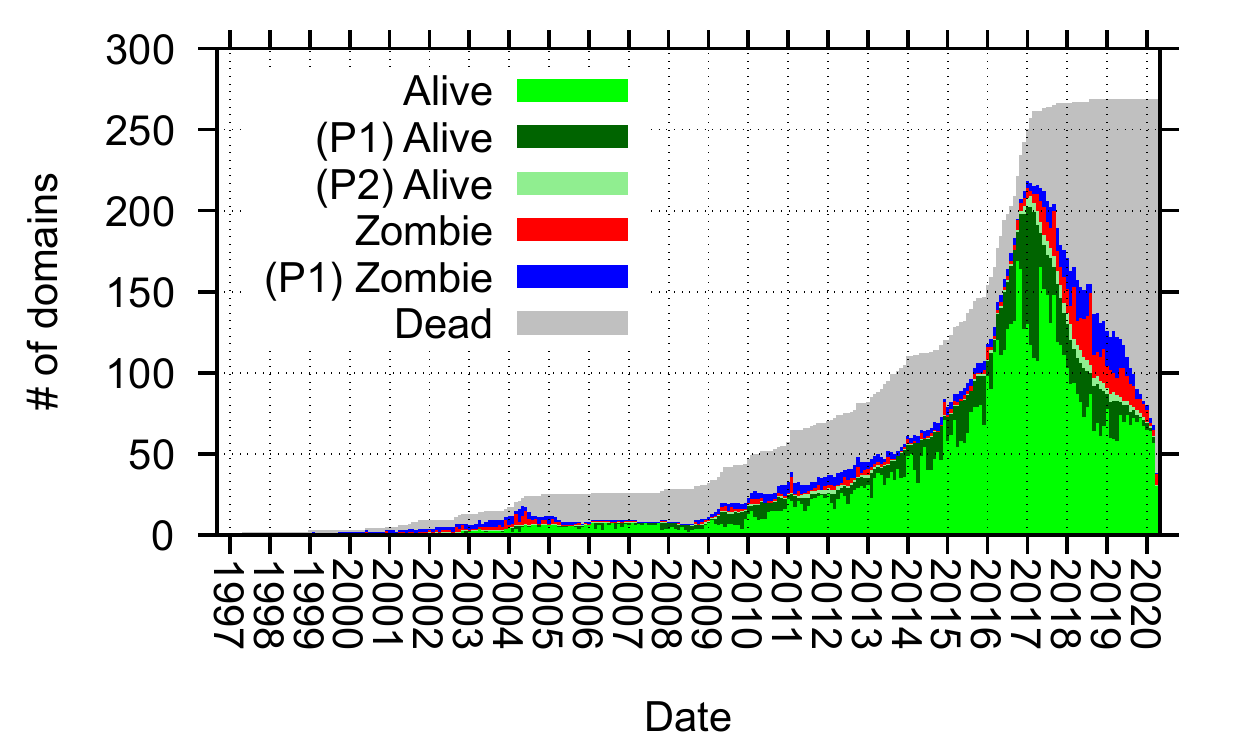}\vspace{-0.3cm}
	    \caption{Histogram of the state of fake news sites for the last 20 years. There is a rise in fake news activity during the USA presidential election years 2016 and 2017, and their sudden fall afterwards.}
	    \label{fig:lifetime_histogram_number}
        \end{minipage}
    \hfill
	\begin{minipage}[t]{0.48\textwidth}
	    \centering
	    \includegraphics[width=1.05\linewidth]{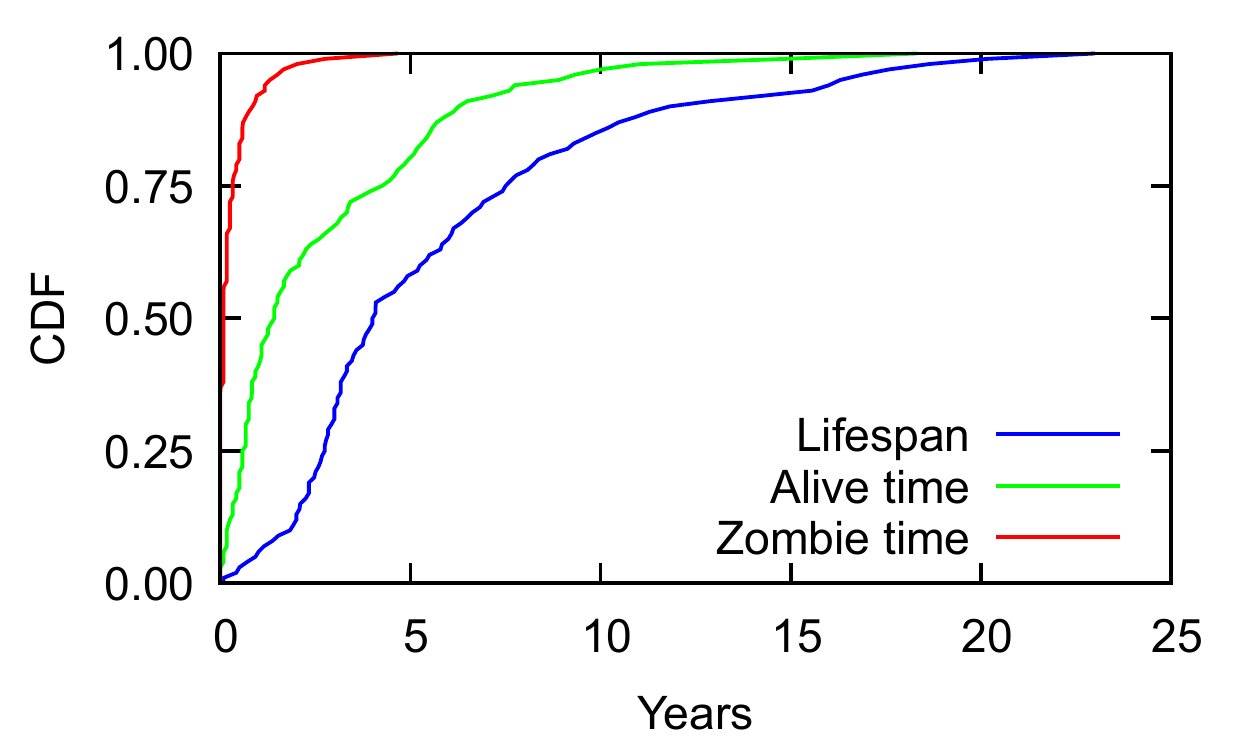}\vspace{-0.3cm}
	    \caption{CDF of lifespan, alive and zombie time for the fake news sites. The median alive and zombie times of fake news sites are as low as 2 and 0.08 years, respectively.}
	    \label{fig:lifetime_life_cdf}
	\end{minipage}\vspace{-0.3cm}
\end{figure}

In Figure~\ref{fig:lifetime_histogram_number}, we present a histogram of the state of the fake news sites in our dataset through our examined period of more than 20 years. In the y-axis, we show the number of domains for each state. In the histogram, we also show the results from the interpolation phases $P1$ and $P2$, and how the plot smooths out, as expected. In this plot, at a given timestamp, we show the number of alive websites (including our interpolation phases), the number of zombie websites, when the remainder from our list is considered dead. Interestingly, we can see the rise in fake news activity during the USA presidential election years 2016 and 2017, and their sudden fall afterwards. During that fall, a portion of these websites turned into zombie state, while the great majority of them  was shut down, especially in the last two years. This can happened either because they fulfilled their purpose (\ie cause polarization or political bias~\cite{10.1145/3366423.3380221}) or after being included in fake news lists and tools for blocking sources of misinformation.

In Figure~\ref{fig:lifetime_life_cdf}, we plot the CDF of the lifespan, alive and zombie time for the fake news sites studied. The lifespan is the absolute maximum that such websites were found to exist on the web. As we can see, their lifespan is found to be about 4 years in median values, when on the other hand, alive and zombie times are lower, with median values of only 2 and 0.08 years, respectively!

\subsection{Do fake news sites synchronize on their uptime and content?}
As a next step, we set out to explore, whether fake news sites appear to synchronize (i) on the times they are available on the Web, and (ii) the content they serve.

\point{Uptime synchronization} To investigate the possible synchronization of their uptime, we assume that each website's sequence of alive or zombie states represents a binary time series and we focus on the last 5 years of the fake news activity (\ie  2015-2020). To retrieve a cleaner signal and differentiate time series that synchronize across websites, we perform an aggregation at the quarter level (\ie 3-month granularity) instead of monthly level.
Thus, the final time series reflects quarters, and each one has 3 possible values (1, 2, or 3) for the number of months from the quarter that alive state was registered.
Then, we compute measures of correlation between pairs of websites, using their quarterly-aggregated time series. For the comparison of time series, we use the pairwise euclidean distance for each pair of fake news sites.

Our method was able to identify several couples and a trio of websites with identical time series for the 2015-2020 timeframe (\ie euclidean distance of zero) categorized into 3 different types:
\begin{enumerate}
	\item $[$the-insider.co, ladylibertynews.com, amposts.com$]$ and $[$dailyinfobox.com, times.com.mx$]$. The websites of this trio and this pair were alive at the same time only for 1/12 quarters of their time series.
	\item $[$coed.com, rickwells.us$]$ The sites of this pair were alive for each of the 12 quarters of the time series.
	\item $[$dailyinfobox.com, times.com.mx$]$, $[$usapolitics24hrs.com, 24wpn.com$]$, $[$politicalo.com, religionlo.com$]$, $[$aurora-news.us, DonaldTrumpPotus45.com$]$, $[$washingtonpost.com.co, drudgereport.com.co$]$, $[$coed.com, rickwells.us$]$ and $[$usaonlinepolitics.com, dailynewsposts.info$]$. These pairs were alive for 2-5 quarters in total.
\end{enumerate}

\point{Content synchronization} To investigate how fake news sites may synchronize their content (in the same time window: 2015-2020), we developed a pipeline to compare pairs of fake news sites with respect to the content they publish.
First, using the Beautiful Soup~\cite{beautifulsoup} library, we extract the text from each website\footnote{There are more modern techniques for article content extraction available such as $newspaper3k$~\cite{newspaper3k} or $readability.js$\cite{readability}, but they are not applicable in our case because they are optimized with heuristics that extract content from full articles rather than landing pages.}. After performing text pre-processing tasks on the extracted content (\ie tokenization, removal of stop-words and lemmatization), we vectorize the documents using a typical TFIDF process~\cite{ramos2003using}. Such vectors were created for each website and timestamp that it had content available. To compare these vectors, we use cosine the similarity metric~\cite{cosine2019} and we set a threshold of 0.5 to select pairs that appear to have high similarity. With this threshold, we ended up with 22 distinct pairs of websites. Upon manually inspecting the pairs at their matched timestamps, we make the following observations regarding fake news site content synchronizations:

\begin{figure}[t]
	\centering
	\begin{minipage}[t]{0.48\textwidth}
	\centering
	\includegraphics[width=1.05\linewidth]{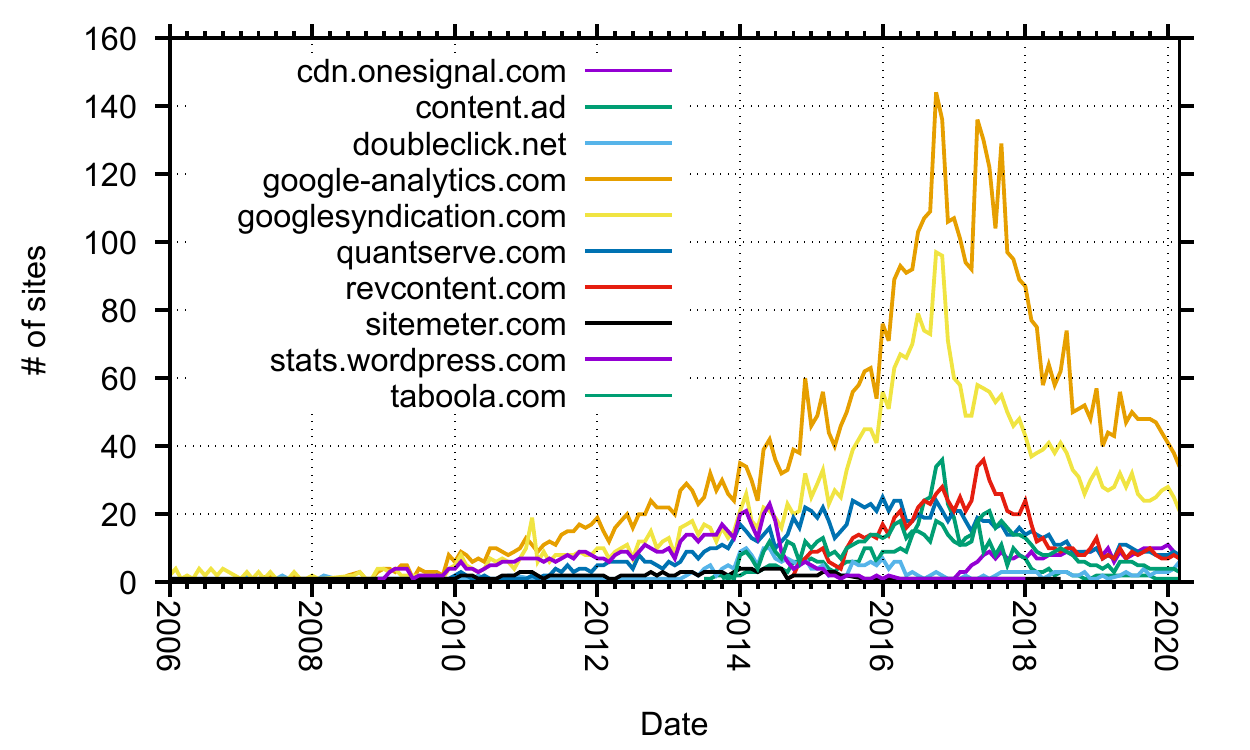}\vspace{-0.3cm}
	\caption{Top 10 third-party domains in fake news sites through time. During the peak of 2016-2017, we see a significant increase in the use of analytics but not a similar increase in the use of ad related third-parties.}
	\label{fig:advertising_top_advertisers}
    \end{minipage}
    \hfill
	\begin{minipage}[t]{0.48\textwidth}
	\centering
		\centering
	\includegraphics[width=1.05\linewidth]{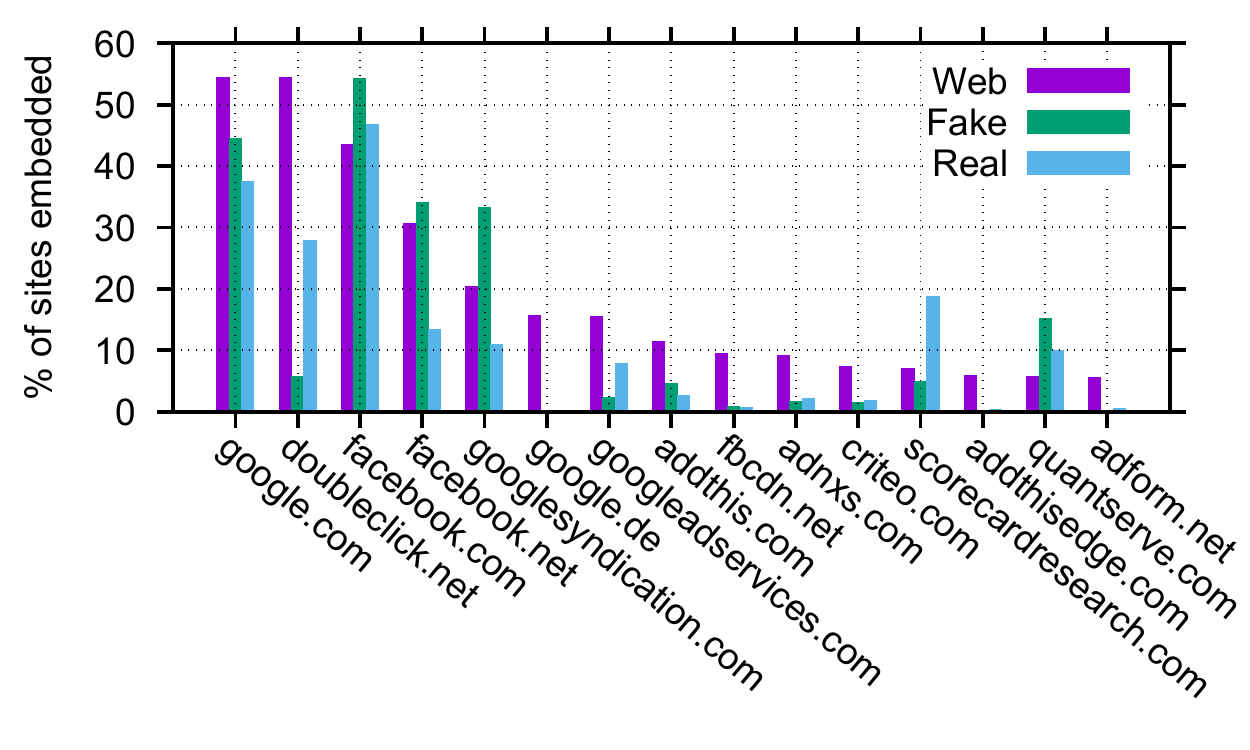}
	\vspace{-0.7cm}
	\caption{Top 15 third-parties on Web, ranked by $Whotracks.me$, and their respective coverage on fake and real news sites of our data.}

	\label{fig:advertising_whotracksme_ads_top15}
	\end{minipage}\vspace{-0.3cm}
\end{figure}

\begin{enumerate}
    \item $[$newslo.com, religionlo.com, politicalo.com, politicops.com, politicot.com$]$. This group of websites consist of 5 domains, with all of them offering the exact same content. The longest period of synchronization was between $[$newslo.com, religionlo.com, politicalo.com$]$, which lasted from 09/2015 to 04/2016. These 7 consecutive months of content synchronization were joined by the pair $[$politicops.com, politicot.com$]$ in the last month.
    \item $[$16wmpo.com, newsdaily12.com, local31news.com$]$. This was an active group for the period between 07/2017 and 11/2017.
    \item $[$usatoday.com.co, washingtonpost.com.co, drudgereport.com.co$]$. This group was synchronized on 07/2015.
\end{enumerate}

We observe that several of the pairs and even portions of the groups above overlap with the uptime synchronization study presented earlier. As a consequence, we believe our proposed methodology of studying synchronization of content and uptime of websites can enable a fake news detection process to select websites that have suspiciously high similarity in their uptime and content, for further examination and even blocking, if needed.

\subsection{Which third-party trackers were persistently embedded in such websites through time?}
Contrary to most popular news sites that progressively move towards paywalling their high quality content~\cite{10.1145/3366423.3380217}, fake news sites rely on ads to make profit. Indeed, some of these websites were even created with the sole purpose of luring ad clicks by publishing clickbait content~\cite{cvetkovska_belford_silverman_feder,macedonian2019}.

To understand which third-party advertising entities provide tracking and other ad-related functionality to fake news sites, we study the third-party domains embedded in these sites through time. Specifically, we parse all collected HTML content per fake news site for each timestamp in our dataset and by using the AdblockPlus blacklist~\cite{adblockplus}, we identify $55$ such third-party domains in the HTML body of at least one website for one timestamp.

In Figure~\ref{fig:advertising_top_advertisers}, we show the number of websites that the top 10 third-party domains were embedded in, per timestamp (\ie month) in the list of fake news sites. These top 10 third-party domains were selected based on their cumulative appearance across fake news sites and across all timestamps. Evidently, analytics and ad entities dominate the top 10 list residing in 91.8\% of all the fake news sites in our dataset. Interestingly, during the aforementioned peak of 2016-2017, we do see a significant increase in the use of analytics (\ie \textit{google-analytics}, and \textit{googlesyndication}) but not an increase in the use of ad related third-parties. This phenomenon shows that, the majority of the marked as fake news sites that were created within this time window (US pre-election period), had purposes other than monetizing their published content (\eg polarize, deliver misinformation, \etc).

\begin{table}[t]
    \centering
       \footnotesize
    \begin{tabular}{lr}
    \toprule
    {\bf Feature}         &   {\bf Range of values for feature} \\ \midrule
        Global rank     &   $[48,\dots,1.2M]$           \\
        Country rank    &   $[3,\dots,816K]$            \\
        Category rank   &   $[1,\dots,19K]$             \\ \hline
        Country (majority of traffic)&   32 countries   \\
        Category        &   42 distinct web categories  \\
        Total visits    &   $[0,\dots,987M]$            \\
        Pages per visit &   $[0,\dots,13.32]$           \\
        Bounce rate     &   $[0\%,\dots,95.37\%]$         \\
        Traffic source direct   &   $[2\%,\dots,100\%]$ \\
        Traffic source referrals&   $[0\%,\dots,69\%]$  \\
        Traffic source search   &   $[0\%,\dots,87\%]$  \\
        Traffic source social   &   $[0\%,\dots,96\%]$  \\
        Traffic source mail     &   $[0\%,\dots,25\%]$  \\
        Traffic source display  &   $[0\%,\dots,95\%]$  \\ \bottomrule
    \end{tabular}
    \caption{Traffic-related features, used for classification.}\vspace{-0.5cm}
    \label{tab:features}
\end{table}

Next, in Figure~\ref{fig:advertising_whotracksme_ads_top15}, we use the data provided by $Whotracks.me$~\cite{whotracksme} to compare the most embedded third-parties on the web, with the ones found in the fake and real news sites of our dataset for the period of 2016-2017. Interestingly, we see Google's \textit{doubleclick} and \textit{googleadservices} residing in less than 6\% and 2\% of the fake news sites, respectively, when they have presence in more than 27\% and 8\% of the real news sites of our dataset, respectively. Similarly, \textit{scorecardresearch} (third biggest web beacons-based tracking service, owned by ComScore~\cite{scorecard}) is present in less than 6\% of the fake news sites but in more than 19\% of the real news sites of our dataset. On the other hand, \textit{facebook} and \textit{quantserve} (second biggest web beacons-based tracking service, owned by Quantcast~\cite{quantserve}) are present in more marked-as-fake news sites than real ones.

%% file: sections/5_ml-classification.tex
\section{Automatic Fake News Classification}
\label{sec:classifier}

Our earlier network traffic analysis of such websites revealed that it is possible for some of these features to be good at distinguishing the nature of the news website, such as number of visits, bounce rate, backlinks, etc.
Thus, we were inspired to build an automated tool that performs the following tasks:
\begin{enumerate}[topsep=0cm]
    \item Retrieves network data for each website from common sources such as \texttt{similarweb} or \texttt{checkpagerank}
    \item Preprocesses the data and extracts related features on network traffic activity
    \item Applies a machine learning (ML) model that classifies the given website as serving fake or real news
\end{enumerate} 

\noindent In order to built this ML classifier, we performed a fresh crawl (February 2021) on our previously mentioned lists of fake and real news websites, and we trained and evaluated our envisioned ML classifier.

\begin{table}[t]
    \centering
    \footnotesize
    \begin{tabular}{llrrrrrr}
    \toprule
    {\bf \#}  &   {\bf Method}                  &   {\bf TP Rate} &   {\bf FP Rate} & {\bf Precision}  &  {\bf Recall}   &   {\bf F1}      &   {\bf AUC}     \\ \midrule
    1   &   Random Forest (RF)      &   0.942   &   0.059   &   0.942   &   0.942   &   0.942   &   0.976   \\
    2   &   Logistic Regression (LR)&   0.915   &   0.091   &   0.916   &   0.915   &   0.915   &   0.968   \\
    3   &   Naive Bayes (NB)        &   0.876   &   0.136   &   0.882   &   0.876   &   0.875   &   0.948   \\
    4   &   Neural Net (15x45x2)    &   0.911   &   0.092   &   0.911   &   0.911   &   0.911   &   0.955   \\ \hline
    5   &   Random Forest (RF):0<Rank<10K       &   0.929   &   0.378   &   0.924   &   0.929   &   0.923   &   0.925   \\
    6   &   Random Forest (RF):10K<Rank<1.3M    &   0.917   &   0.096   &   0.917   &   0.917   &   0.917   &   0.970   \\
    7   &   \#6 classifier tested on data of \#5&   0.923   &   0.174   &   0.932   &   0.923   &   0.926   &   0.942   \\
    \bottomrule
    \end{tabular}
    \caption{Performance metrics from ML binary classification of websites as showing fake or real news.}
    \label{tab:classification-results}\vspace{-0.5cm}
\end{table}

Based on previously mentioned network traffic metrics (summarized in Table~\ref{tab:features}) we train different ML classifiers for automatic classification of news websites as ``real'' or ``fake''.
As a basic preprocessing step, we removed features with very little to zero variability.
Our dataset for training and testing is fairly balanced, with ``real'' news websites being 278 and ``fake'' being 239.
The difference from the previous numbers lies in the fact that we did extra steps for removing websites that did not have scores across all metrics.
We applied 10-fold cross-validation on the available data, and trained and tested various techniques.
We measured standard ML performance metrics such as True Positive and False Positive Rates, Precision and Recall, F1 score and Area Under the Receiver Operating Curve (AUC).
The scores were weighted to take into account individual performance metrics per class weight.

Table~\ref{tab:classification-results} shows the results achieved with different basic classifiers when all the dataset is used (upper part, classifiers \#1-\#4).
We find that the typical Random Forest classifier performs very well across the board, with high True Positive and low False Positive rates, and higher Precision and Recall than the other ML methods.

Given that the amount of traffic and other features used here are naturally correlated with each other (\eg~a highly ranked website should attract more visits, \etc), we also test the scenario where we split our dataset into two major groups of ranked websites (highly popular with rank <10K, and rest), to check if the ML classification is still possible under similarly ranked websites.
The results, shown in Table~\ref{tab:classification-results} (lower part, classifiers \#5 and \#6) demonstrate that it is possible to achieve very good performance, even when controlling for the rank of websites.
Furthermore, classifier row \#7 checks the scenario where data from the lower ranked websites (\ie~rank>10K) are used to train a classifier that is then tested on data from higher ranked websites (\ie~rank<10K).
Interestingly, the performance still remains high, showing that examples of fake news websites from lower ranking can be useful to distinguish such websites even at higher ranks.

%% file: sections/6_related.tex
\section{Related Work}
\label{sec:related}
Numerous studies are attempting to explore the characteristics of fake news and its spread. In~\cite{fourney2017geographic}, authors conducted a traffic analysis to websites known for publishing fake news in the months preceding the 2016 US presidential election. Although the study also includes features as traffic sources and temporal trends, our work significantly diverges from that. We analyze a much greater set of websites, we compare fake with real news websites and we do not focus on social networks or the elections. In ~\cite{vargo2018agenda} authors perform a 3-year long study on fake news websites prior (2014-2016). Their analysis includes time-series modelling for causality testing. This is one of the few studies that are including time-series analysis. However, the method differs from ours in substantial manners.

As important as understanding fake news dynamics is, we cannot avoid mentioning the significant efforts to identify and flag fake news stories. Based on a survey~\cite{10.1145/3305260}, fake news can be identified with content-based, feedback-based and intervention-based methods. In~\cite{shu2017mining} authors characterize detection as knowledge-based, stance-based, style-based and propagation-based. While there is a large number of publications in the Fake News detection area, we will only give specific examples. Check-it~\cite{paschalides2019check} is an ensemble method that combines different signals to generate a flag about a news article / social media post. The aforementioned signals are the domain name, linguistic features, reputation score and others. NELA~\cite{horne2018assessing} creates and combines credibility scores of the news article and the news source. Although it is possible to combine different methods to solve this problem, most papers focus on a single - most narrowed down - approach. For example, in a different publication, Shu~\cite{shu2019context} also argues about the role of social context for fake news detection. 

In~\cite{zannettou2019web}, authors provide a comprehensive overview of existing research on the false information ecosystem. In~\cite{Ghanem2019AnEA} authors show that fake news aim to affect the emotions of the readers, and ultimately deceive them. The authors create a neural network capable of detecting false news from such an effect. In \cite{fourney2017geographic}, it is clear that aggregate voting patterns were strongly correlated with the average daily fraction of users visiting websites serving fake news. In \cite{bovet2019influence}, authors dive into the dynamics and influence of fake news on Twitter during the 2016 US presidential election. With a dataset of 30 million tweets and the opensources.co list, it finds that 25\% of these tweets spread either fake or extremely biased news. Based on \cite{10.1145/3131365.3131390}, the online social network ecosystems seem to interplay, and information shared in a network is affecting the information flow in another.

A lot of work has been done on Twitter disinformation. In~\cite{Jin2017RumorDO}, authors present a thorough analysis of rumour tweets from the followers of two presidential candidates. It is also shown by \cite{zannettou2019disinformation} that many trolls have been sponsored externally for ultimate goals. \cite{shao2017spread} proves that bots play a key role in the Twitter misinformation ecosystem, targeting influential users for misinformation spreading. As shown by \cite{grinberg2019fake}, the vast majority of fake news is spread by an extremely small number of sources, forming clusters. This study sheds light on the target groups as well, being conservative-leaning, older and highly engaged with political news individuals. Following a different trajectory, \cite{7589045} points to images as being very crucial content for the news verification process.

%% file: sections/7_conclusion.tex
\section{Summary \& Conclusion}

In this paper, we performed a first of its kind investigation on the fake news ecosystem.
We studied and compared their user engagement by analyzing traffic-related metrics for fake and real news websites. Additionally, we explored the lifetime of fake news sites and their typical uptime periods over a time range of more than 20 years, and we proposed a methodology to study how websites may be synchronizing their uptime periods, and the content they serve during these uptime periods. Our findings can be summarized as follows:
\begin{itemize}[leftmargin=0.5cm]
    \item The median real news site tends to have a larger number of pages (\ie 2.18 pages, on average) visited per user than the median fake news site (\ie 1.72 pages, on average).
    \item On average, visits last longer (\ie 198.8 seconds) in real news sites, than in fake news websites (\ie 163.4 seconds).
    \item The median fake news site is being accessed mostly directly, when corresponding real news site via search engines.
    \item The median real news site has a significantly lower bounce rate, compared to the median fake news one.
    \item The median fake news site in our dataset scores 4.7K backlinks, when the median real news website more than 23.2M backlinks! Fake news sites have lower portions of EDU backlinks and referrals, as well as GOV backlinks than the real news sites.
    \item Fake news sites have about 2 orders of magnitude lower number of referring domains than real news.
    \item The median alive and zombie times of fake news sites are as low as 2 and 0.08 years, respectively.
    \item There was a significant rise in fake news activity during the USA presidential election years 2016-2017, and there was a rapid fall afterwards.
    \item We detect numerous clusters of websites synchronizing their uptime and content for long periods of time. 
    \item During this period, we see a significant increase in the use of analytics but not an increase in the use of ad related third-parties from the fake news sites. This shows that the majority of the fake news sites created within this time window, had purposes other than monetizing their published content (\eg polarize, deliver misinformation, \etc).
    \item Domains like \textit{doubleblick}, \textit{googleadservices} and \textit{scorecardresearch} tend  to have higher presence in real news sites than in marked-as-fake ones. On the contrary, \textit{facebook} and \textit{quantserve} have higher presence in fake news sites.
\end{itemize}

Our findings enabled us to characterize the traffic and behavior of fake news sites, and build a novel, content-agnostic machine learning (ML) classifier for automatic detection of fake news websites that are not yet included in manually curated blacklists. We tested various supervised and unsupervised ML methods for our classifier which achieved a very good performance: \accuracy.

In the future, we plan to investigate how such ML model can be updated at pseudo-real time, with data collection that happens at regular intervals or at the discovery of a news website.
This effort can be done in a crowd-sourced fashion across multiple online users, by deploying the ML pipeline envisioned earlier into a browser plugin.
Then, the plugin can 1) perform the crawling of network metadata for the website visited from its user, and 2) apply the ML model we provide.
The plugin can also report these metadata per website to a centralized location for updating our ML model.
In case privacy of users is at stake, privacy-preserving methodologies can be used, that employ Federated Learning techniques for training the ML model, coupled with local differential privacy applied at the user devices.

%% file: sections/8_acks.tex
\section{Acknowledgments}
The research leading to these results received funding from the EU H2020 Research and Innovation programme under grant agreements No 830927 (Concordia), No 871793 (Accordion), No 871370 (Pimcity) and Marie Sklodowska-Curie No 690972 (Protasis).
These  results reflect only the authors' view and the Commission is not responsible for any use that may be made of the information it contains.

%% file: paper.bbl
\begin{thebibliography}{10}

\bibitem{fakeNewsVSrealNews}
Claire~Pedersen Juju~Chang, Jake~Lefferman and Geoff Martz.
\newblock When fake news stories make real news headlines.
\newblock
  \url{https://www.theatlantic.com/ideas/archive/2019/06/fake-news-republicans-democrats/591211/},
  2016.

\bibitem{fakeNewsSurvey}
David~A. Graham.
\newblock Some real news about fake news.
\newblock
  \url{https://www.theatlantic.com/ideas/archive/2019/06/fake-news-republicans-democrats/591211/},
  2019.

\bibitem{zannettou2019web}
Savvas Zannettou, Michael Sirivianos, Jeremy Blackburn, and Nicolas Kourtellis.
\newblock The web of false information: Rumors, fake news, hoaxes, clickbait,
  and various other shenanigans.
\newblock {\em Journal of Data and Information Quality (JDIQ)}, 11(3):1--37,
  2019.

\bibitem{Ghanem2019AnEA}
Bilal Ghanem, Paolo Rosso, and Francisco M.~Rangel Pardo.
\newblock An emotional analysis of false information in social media and news
  articles.
\newblock {\em ACM Transactions on Internet Technology (TOIT)}, 20:1 -- 18,
  2019.

\bibitem{7589045}
Z.~{Jin}, J.~{Cao}, Y.~{Zhang}, J.~{Zhou}, and Q.~{Tian}.
\newblock Novel visual and statistical image features for microblogs news
  verification.
\newblock {\em IEEE Transactions on Multimedia}, 19(3):598--608, 2017.

\bibitem{10.1145/3131365.3131390}
Savvas Zannettou, Tristan Caulfield, Emiliano De~Cristofaro, Nicolas
  Kourtellis, Ilias Leontiadis, Michael Sirivianos, Gianluca Stringhini, and
  Jeremy Blackburn.
\newblock The web centipede: Understanding how web communities influence each
  other through the lens of mainstream and alternative news sources.
\newblock In {\em Proceedings of the 2017 Internet Measurement Conference}, IMC
  ’17, page 405–417, New York, NY, USA, 2017. Association for Computing
  Machinery.

\bibitem{Jin2017RumorDO}
Zhiwei Jin, Juan Cao, Han Guo, Yongdong Zhang, Yu~Wang, and Jiebo Luo.
\newblock Rumor detection on twitter pertaining to the 2016 u.s. presidential
  election.
\newblock {\em ArXiv}, abs/1701.06250, 2017.

\bibitem{10.1145/3305260}
Karishma Sharma, Feng Qian, He~Jiang, Natali Ruchansky, Ming Zhang, and Yan
  Liu.
\newblock Combating fake news: A survey on identification and mitigation
  techniques.
\newblock {\em ACM Trans. Intell. Syst. Technol.}, 10(3), April 2019.

\bibitem{zannettou2019disinformation}
Savvas Zannettou, Tristan Caulfield, Emiliano De~Cristofaro, Michael
  Sirivianos, Gianluca Stringhini, and Jeremy Blackburn.
\newblock Disinformation warfare: Understanding state-sponsored trolls on
  twitter and their influence on the web.
\newblock In {\em Companion Proceedings of The 2019 World Wide Web Conference},
  pages 218--226, 2019.

\bibitem{grinberg2019fake}
Nir Grinberg, Kenneth Joseph, Lisa Friedland, Briony Swire-Thompson, and David
  Lazer.
\newblock Fake news on twitter during the 2016 us presidential election.
\newblock {\em Science}, 363(6425):374--378, 2019.

\bibitem{lazer2018science}
David~MJ Lazer, Matthew~A Baum, Yochai Benkler, Adam~J Berinsky, Kelly~M
  Greenhill, Filippo Menczer, Miriam~J Metzger, Brendan Nyhan, Gordon
  Pennycook, David Rothschild, et~al.
\newblock The science of fake news.
\newblock {\em Science}, 359(6380):1094--1096, 2018.

\bibitem{shao2017spread}
Chengcheng Shao, Giovanni~Luca Ciampaglia, Onur Varol, Alessandro Flammini, and
  Filippo Menczer.
\newblock The spread of fake news by social bots.
\newblock {\em arXiv preprint arXiv:1707.07592}, 96:104, 2017.

\bibitem{fourney2017geographic}
Adam Fourney, Miklos~Z Racz, Gireeja Ranade, Markus Mobius, and Eric Horvitz.
\newblock Geographic and temporal trends in fake news consumption during the
  2016 us presidential election.
\newblock In {\em CIKM}, volume~17, pages 6--10, 2017.

\bibitem{similarweb}
Nir~Cohen Or~Offer.
\newblock Similarweb: Website traffic statistics \& analytics.
\newblock \url{www.larweb.com}, 2007.

\bibitem{checkpagerank}
{Caye Group}.
\newblock Check page rank - check your pagerank free!
\newblock \url{www.checkpagerank.net}, 2004.

\bibitem{OpenSources2017}
OpenSources.
\newblock
  \url{https://github.com/BigMcLargeHuge/opensources/blob/master/sources/sources.csv},
  2017.

\bibitem{bovet2019influence}
Alexandre Bovet and Hern{\'a}n~A Makse.
\newblock Influence of fake news in twitter during the 2016 us presidential
  election.
\newblock {\em Nature communications}, 10(1):1--14, 2019.

\bibitem{horne2018assessing}
Benjamin~D Horne, William Dron, Sara Khedr, and Sibel Adali.
\newblock Assessing the news landscape: A multi-module toolkit for evaluating
  the credibility of news.
\newblock In {\em Companion Proceedings of the The Web Conference 2018}, pages
  235--238, 2018.

\bibitem{bsdetector}
Daniel Sieradski.
\newblock B.s. detector.
\newblock \url{https://gitlab.com/bs-detector/bs-detector}, 2017.

\bibitem{wayback}
{Internet Archive}.
\newblock Wayback machine.
\newblock \url{https://archive.org/web/}, 2001.

\bibitem{puppeteer}
{Puppeteer Developers}.
\newblock Puppeteer: Headless chrome node.js api.
\newblock \url{https://pptr.dev/}, 2020.

\bibitem{similarweb-average-visit}
SimilarWeb.
\newblock Similar web average visit.
\newblock
  \url{https://support.similarweb.com/hc/en-us/articles/115000501485-Average-Visit-Duration}.

\bibitem{similarweb-bouncerate}
SimilarWeb.
\newblock Similar web bound rate.
\newblock
  \url{https://support.similarweb.com/hc/en-us/articles/115000501625-Bounce-Rate}.

\bibitem{10.1145/3366423.3380221}
Pushkal Agarwal, Sagar Joglekar, Panagiotis Papadopoulos, Nishanth Sastry, and
  Nicolas Kourtellis.
\newblock Stop tracking me bro! differential tracking of user demographics on
  hyper-partisan websites.
\newblock In {\em Proceedings of The Web Conference 2020}, WWW '20, page
  1479–1490, New York, NY, USA, 2020. Association for Computing Machinery.

\bibitem{beautifulsoup}
Leonard Richardson.
\newblock Beautiful soup documentation.
\newblock \url{https://www.crummy.com/software/BeautifulSoup/bs4/doc/}, 2020.

\bibitem{newspaper3k}
Lucas Ou-Yang.
\newblock Newspaper3k: Article scraping \& curation.
\newblock \url{https://newspaper.readthedocs.io/en/latest/}, 2013.

\bibitem{readability}
{Mozilla Foundation}.
\newblock Readability.js: A standalone version of the readability library used
  for firefox reader view.
\newblock \url{https://github.com/mozilla/readability}, 2020.

\bibitem{ramos2003using}
Juan Ramos et~al.
\newblock Using tf-idf to determine word relevance in document queries.
\newblock In {\em Proceedings of the first instructional conference on machine
  learning}, volume 242, pages 133--142. Piscataway, NJ, 2003.

\bibitem{cosine2019}
Cosine similarity, May 2019.

\bibitem{10.1145/3366423.3380217}
Panagiotis Papadopoulos, Peter Snyder, Dimitrios Athanasakis, and Benjamin
  Livshits.
\newblock Keeping out the masses: Understanding the popularity and implications
  of internet paywalls.
\newblock In {\em Proceedings of The Web Conference 2020}, WWW '20, page
  1433–1444, New York, NY, USA, 2020. Association for Computing Machinery.

\bibitem{cvetkovska_belford_silverman_feder}
Saska Cvetkovska, Aubrey Belford, Craig Silverman, and J.~Lester Feder.
\newblock The secret players behind macedonia's fake news sites.
\newblock
  \url{https://www.occrp.org/en/spooksandspin/the-secret-players-behind-macedonias-fake-news-sites},
  2018.

\bibitem{macedonian2019}
BBC Future.
\newblock I was a macedonian fake news writer.
\newblock
  \url{https://www.bbc.com/future/article/20190528-i-was-a-macedonian-fake-news-writer},
  May 2019.

\bibitem{adblockplus}
{fanboy, MonztA, Famlam, Khrin}.
\newblock Easylist filterlist project.
\newblock \url{https://easylist.to/pages/about.html}, 2020.

\bibitem{whotracksme}
Arjaldo Karaj, Sam Macbeth, Rémi Berson, and Josep~M. Pujol.
\newblock Whotracks.me: Shedding light on the opaque world of online tracking,
  2018.

\bibitem{scorecard}
Joanna~Geary Teodora~Beleaga.
\newblock Scorecardresearch (comscore): What is it and what does it do?
\newblock
  \url{https://www.theguardian.com/technology/2012/apr/23/scorecardresearch-tracking-trackers-cookies-web-monitoring},
  2012.

\bibitem{quantserve}
Joanna~Geary James~Ball.
\newblock Quantserve (quantcast): What is it and what does it do?
\newblock
  \url{https://www.theguardian.com/technology/2012/apr/23/quantcast-tracking-trackers-cookies-web-monitoring},
  2012.

\bibitem{vargo2018agenda}
Chris~J Vargo, Lei Guo, and Michelle~A Amazeen.
\newblock The agenda-setting power of fake news: A big data analysis of the
  online media landscape from 2014 to 2016.
\newblock {\em New media \& society}, 20(5):2028--2049, 2018.

\bibitem{shu2017mining}
Kai Shu, Amy Sliva, Suhang Wang, Jiliang Tang, and Huan Liu.
\newblock Fake news detection on social media: A data mining perspective.
\newblock {\em SIGKDD Explor. Newsl.}, 19(1):22–36, September 2017.

\bibitem{paschalides2019check}
Demetris Paschalides, Chrysovalantis Christodoulou, Rafael Andreou, George
  Pallis, Marios~D Dikaiakos, Alexandros Kornilakis, and Evangelos Markatos.
\newblock Check-it: A plugin for detecting and reducing the spread of fake news
  and misinformation on the web.
\newblock In {\em 2019 IEEE/WIC/ACM International Conference on Web
  Intelligence (WI)}, pages 298--302. IEEE, 2019.

\bibitem{shu2019context}
Kai Shu, Suhang Wang, and Huan Liu.
\newblock Beyond news contents: The role of social context for fake news
  detection.
\newblock In {\em Proceedings of the Twelfth ACM International Conference on
  Web Search and Data Mining}, WSDM '19, page 312–320, New York, NY, USA,
  2019. Association for Computing Machinery.

\end{thebibliography}
